\documentclass[prl,twocolumn,floatfix,superscriptaddress]{revtex4-2}

\usepackage{amssymb}
\usepackage{amsmath}       
\usepackage{graphicx}
\usepackage{hyperref}
\usepackage{color}
\usepackage{enumerate}
\usepackage[normalem]{ulem}

\begin{document}

\clubpenalty=10000
\widowpenalty=10000
\brokenpenalty=10000
\tolerance=9000
\hyphenpenalty=10000

\title{\boldmath Spin-lattice and magnetoelectric
couplings enhanced by orbital degrees of freedom in polar magnets \unboldmath}

\author{Vilmos~Kocsis}
\affiliation{RIKEN Center for Emergent Matter Science (CEMS), Wako, Saitama 351-0198, Japan}
\affiliation{Institut f\"ur Festk\"orperforschung, Leibniz IFW-Dresden, 01069 Dresden, Germany}

\author{Yusuke~Tokunaga}
\affiliation{RIKEN Center for Emergent Matter Science (CEMS), Wako, Saitama 351-0198, Japan}
\affiliation{Department of Advanced Materials Science, University of Tokyo, Kashiwa 277-8561, Japan}

\author{Toomas~R{\~o}{\~o}m}
\affiliation{National Institute of Chemical Physics and Biophysics,
12618 Tallinn, Estonia}

\author{Urmas~Nagel}
\affiliation{National Institute of Chemical Physics and Biophysics,
12618 Tallinn, Estonia}

\author{Jun~Fujioka}
\affiliation{Institute of Materials Science, University of Tsukuba, Ibaraki 305-8573, Japan}

\author{Yasujiro~Taguchi}
\affiliation{RIKEN Center for Emergent Matter Science (CEMS), Wako, Saitama 351-0198, Japan}

\author{Yoshinori Tokura}
\affiliation{RIKEN Center for Emergent Matter Science (CEMS), Wako, Saitama 351-0198, Japan}
\affiliation{Tokyo College and Department of Applied Physics, University of Tokyo, Hongo, Tokyo 113-8656, Japan}

\author{S\'andor~Bord\'acs}
\affiliation{Department of Physics, Institute of Physics, Budapest University of Technology and Economics, M\H{u}egyetem rkp. 3., H-1111 Budapest, Hungary}
\affiliation{Quantum Phase Electronics Center and Department of Applied Physics, University of Tokyo, Tokyo 113-8656, Japan}

\begin{abstract}
Orbital degrees of freedom mediating an interaction between spin and lattice were predicted to raise strong magnetoelectric effect, i.e. realize an efficient coupling between magnetic and ferroelectric orders.
However, the effect of orbital fluctuations have been considered only in a few magnetoelectric materials, as orbital degeneracy driven Jahn-Teller effect rarely couples to polarization.
Here, we explore the spin-lattice coupling in multiferroic Swedenborgites with mixed valence and Jahn-Teller active transition metal ions on a stacked triangular/Kagome lattice using infrared and dielectric spectroscopy.
On one hand, in CaBa$M_4$O$_7$ ($M$ = Co, Fe), we observe strong magnetic order induced shift in the phonon frequencies and a corresponding large change in the dielectric response.
Remarkably, as an unusual manifestation of the spin-phonon coupling, the spin-fluctuations reduce the phonon life-time by an order of magnitude at the magnetic phase transitions.
On the other hand, lattice vibrations, dielectric response, and electric polarization show no variation at the N\'eel temperature of CaBaFe$_2$Co$_2$O$_7$, which is built up by orbital singlet ions.
Our results provide a showcase for orbital degrees of freedom enhanced magnetoelectric coupling via the example of Swedenborgites.
\end{abstract}

\maketitle


Spin-orbit coupling (SOC) is considered among the most essential interactions in condensed matter science, standing in the background of topological insulators~\cite{Hasan2010RMP} and superconductors~\cite{Qi2011RMP}, Dirac and Weyl semimetals~\cite{Liu2014Sci,Xu2015Sci}, Kitaev physics~\cite{Kitaev2006} as well of multiferroics~\cite{Fiebig2005,Tokura2014}.
In the latter compounds, SOC induces magnetoelectric (ME) coupling between electric polarization and magnetism making them interesting for basic research and appealing for applications, however, this interaction is usually weak due to its relativistic nature.~\cite{Katsura2005,Sergienko2006,Jia2006,Jia2007,Arima2007}.
While the relativistic spin-orbit interaction enables the ME coupling on a single (a pair) of magnetic ion(s), theoretical works proposed early that the charge and orbital degrees of freedoms can mediate an enhanced ME interaction via the Kugel-Khomski\u{\i}-type spin-orbital coupling~\cite{Kugel1982,Tokura2000,Brink2008JPCM,Yamauchi2012PRB}.
However, materials realizing this scenario are exceptional, as charge and orbital order alone rarely break the inversion symmetry~\cite{Rado1975PRB,Kato1982JPSJ,Alexe2009AM,Yamauchi2012PRB,Groot2012PRL,Johnson2012PRL,Perks2012NatComm}.
The two most studied cases are Fe$_3$O$_4$, where the ME effect is attributed to the charge and orbital orderings~\cite{Rado1975PRB,Kato1982JPSJ,Alexe2009AM,Yamauchi2012PRB}, and LuFe$_2$O$_4$ in which the ferroelectricity is debated to emerge from charge ordering~\cite{Groot2012PRL}.
Recently, CaMn$_7$O$_{12}$ was also identified with a chiral magnetic structure stabilized by the charge and orbital ordering~\cite{Johnson2012PRL,Perks2012NatComm}.

\begin{center}
    \begin{figure}
 
    \includegraphics[width=8.5truecm]{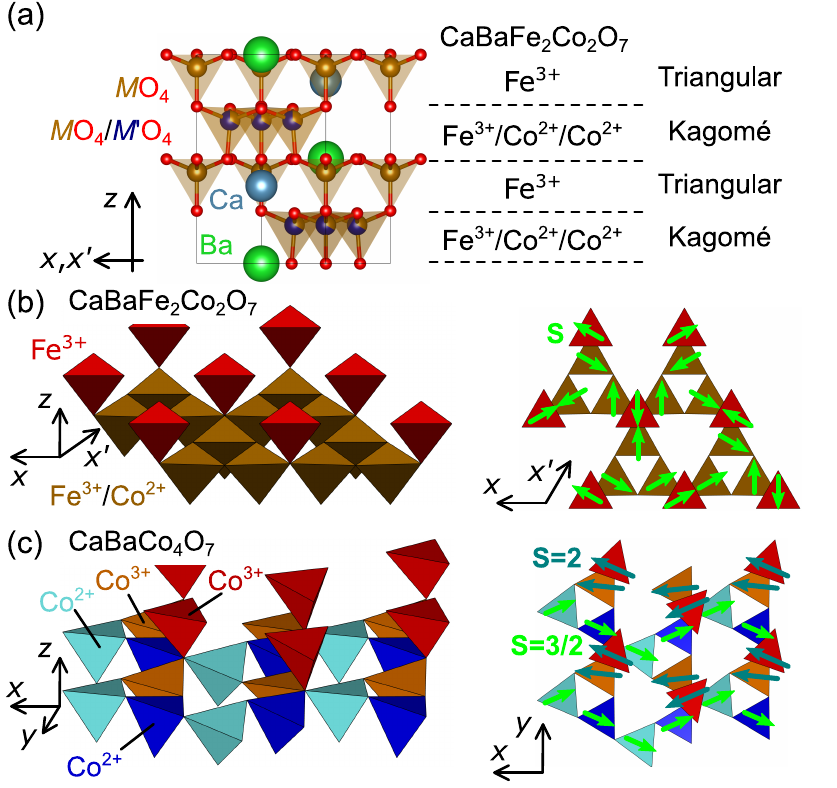}
    \caption{
    (a) The polar structural unit cell of trigonal Swedenborgites are built up by alternating triangular and Kagom\'e layers of co-aligned $M$O$_4$ tetrahedra.
    (b) In the trigonal CaBaFe$_2$Co$_2$O$_7$, one Fe$^{3+}$ ion occupies the triangular lattice, while the remaining Fe$^{3+}$/Co$^{2+}$/Co$^{2+}$ ions are distributed randomly on the Kagome lattice. The $\sqrt{3}\,\times\,\sqrt{3}$-type antiferromagnetic order develops below $T_{\rm N}$=152\,K (spin $\mathbf{S}$, green arrow, reproduced after Ref.~\onlinecite{Reim2018PRB}.)
    (c) The orthorhombic CaBaCo$_4$O$_7$ has charge order and a ferrimagnetic order below $T_{\rm C}$=60\,K, reproduced after Ref.~\onlinecite{Chatterjee2011,Caignaert2010}.}
    \label{Swed_00}
    \end{figure}
    \end{center}

\begin{center}
    \begin{figure*}
 
    \includegraphics[width=17truecm]{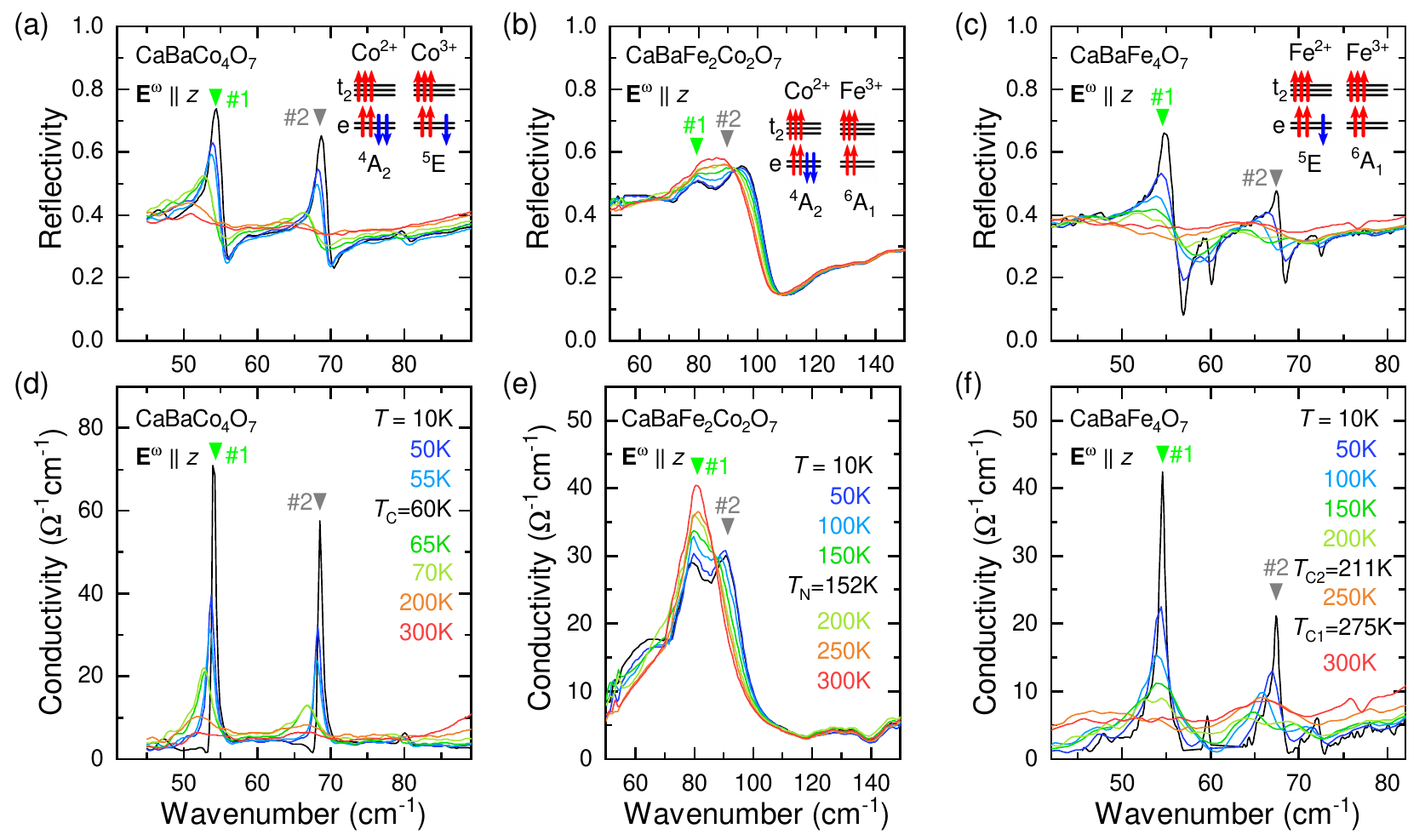}
    \caption{The reflectivity and the optical conductivity spectra of (a,d) CaBaCo$_4$O$_7$, (b,e) CaBaFe$_2$Co$_2$O$_7$, and (c,f) CaBaFe$_4$O$_7$ at selected temperatures in the frequency range of the lowest energy phonon modes.}
    \label{Swed_01}
    \end{figure*}
    \end{center}

Swedenborgites CaBa$M_4$O$_7$ ($M$=Co, Fe) provide another platform to study the interplay between spins and orbitals, but there, unlike the previous examples, the charge degree of freedom is quenched.
The polar Swedenborgites are built up by alternating layers of triangular and kagom\'e sheets of $M$O$_4$ tetrahedra, all pointing to the $c$ axis, as shown in Fig.~\ref{Swed_00}(a).
The $M^{2.5+}$ nominal valence, suggests a 1:1 mixture of $M^{2+}$ and $M^{3+}$ ions, subjected to geometric frustration.
The buckling of the kagom\'e lattice releases the frustration and reduces the symmetry to orthorhombic at $T_{\rm S}$=450\,K~\cite{Caignaert2010,Kocsis2020PRB2SM} and $T_{\rm S}$=380\,K~\cite{Hollmann2011,Kocsis2016PRB} in CaBaCo$_4$O$_7$ and CaBaFe$_4$O$_7$, respectively.
In both compounds, X-ray spectroscopy studies confirmed the coexistence of distinct valences, $M^{2+}$ and $M^{3+}$ (electron configurations sketched in Fig.~\ref{Swed_01}), and suggested charge order with the $M^{3+}$ ions occupying the triangular and one of the kagom\'e sites~\cite{Caignaert2010,Chatterjee2011,Hollmann2011,Cuartero2018IC,Galakhov2018JETP}.
Therefore, both CaBaCo$_4$O$_7$ and CaBaFe$_4$O$_7$ contain the Jahn-Teller active Co$^{3+}$ and Fe$^{2+}$ ions, respectively, though, no further information is available on orbital ordering.
However, the solid-solution CaBaFe$_2$Co$_2$O$_7$ lacks orbital degeneracy, namely solely the orbital singlet Fe$^{3+}$ and Co$^{2+}$ charge states are present in this compound~\cite{Reim2014,Cuartero2018IC,Galakhov2018JETP}.

In CaBaCo$_4$O$_7$, spins order antiferromegnetically at $T_{\rm N}$=70\,K~\cite{Omi2021PRB}, and then a ferrimagnetic structure emerges below $T_{\rm C}$=60\,K~\cite{Caignaert2010,Bordacs2015PRB}, as shown in Fig.~\ref{Swed_00}(c).
The latter phase is accompanied by one of the largest magnetic-order-induced polarization detected so far~\cite{Caignaert2013,Johnson2014} as well as exceptionally large magnetostriction~\cite{Chai2021PRB}.
Its sister compound, CaBaFe$_4$O$_7$ also show peculiar ME properties. It becomes multiferroic close to room temperature, $T_{\rm C1}$=275\,K upon a ferrimagnetic ordering, which is followed by a reorientation transition below $T_{\rm C2}$=211\,K~\cite{Hollmann2011,Kocsis2016PRB}.
CaBaFe$_2$Co$_2$O$_7$ develops an antiferromagnetic structure at $T_{\rm N}$=152\,K~\cite{Soda2006,Reim2018PRB,Reim2014} (Fig.~\ref{Swed_00}(b)), however, its ME properties have been unknown.

In this Letter, we investigate the effect of magnetic ordering on the charge dynamics of Swedenborgites via infrared and dielectric spectroscopy.
We compared members of the material family with and without orbital degree of freedom, and found a strong spin-lattice coupling only in CaBa$M_4$O$_7$ ($M$ = Co, Fe) with Jahn-Teller active ions.
In these pristine compounds, the phonon frequencies show a sudden shift at $T_{\rm C}$, related to the large magnetic-order-induced polarization and magnetocapacitance.
Moreover, we observed an order of magnitude decrease of the phonon life-times at the ferrimagnetic phase transitions.
In contrast, we found no phonon nor dielectric anomalies and negligible change in the pyroelectric polarization upon the magnetic ordering in the orbital-singlet CaBaCo$_2$Fe$_2$O$_7$.
Therefore, our results highlight the importance of orbital degrees of freedom in the enhancement of the spin-lattice interaction and the ME effect in multiferroics.


Large single crystals of CaBaCo$_4$O$_7$, CaBaFe$_4$O$_7$, CaBaFe$_2$Co$_2$O$_7$, and YBaCo$_3$AlO$_7$ were grown by the floating zone technique~\cite{Valldor2008,Valldor2011a,Caignaert2013,Reim2014,Kocsis2016PRB}.
Polarized, near normal incidence reflectivity was measured on polished cuts.
Temperature dependent experiments were carried out up to 40000 cm$^{-1}$ with an FT-IR spectrometer (Vertex80v, Bruker) and a grating-monochromator spectrometer (MSV-370YK, Jasco).
The reflectivity spectrum of each compound was measured up to 250000 cm$^{-1}$ at room temperature with use of synchrotron radiation at UVSOR Institute for Molecular Science, Okazaki, Japan.
The optical conductivity was calculated using the Kramers-Kronig transformation~\cite{Kocsis2020PRB2SM}.
The pyroelectric polarization was obtained by measuring and integrating the displacement current with an electrometer (6517A, Keithley) while the temperature was swept in a Physical Property Measurement System (PPMS, Quantum Design).
The dielectric properties were also measured in a PPMS, using an LCR meter (E4980A, Keysight Technologies) while the ac magnetization was measured in a Magnetic Property Measurement System (MPMS3, Quantum Design).

For quantitative analysis, we fitted the real part of the optical conductivity as a sum of Lorentz oscillators:
\begin{equation}
\sigma \left( \omega \right) =-i \omega\epsilon_0\left[\epsilon_\infty+\sum_{j}\frac{S_j}{\omega_{0,j}^2-\omega^2-i\gamma_j\omega}\right],
\end{equation}
where $\omega_{0,j}$, $S_j$, and $\gamma_j$ are the frequency, oscillator strength, and damping rate of the $j^{th}$ mode, and $\epsilon_\infty$ is the high-frequency dielectric constant, respectively.


In Fig.~\ref{Swed_01}, we show the temperature dependence of the reflectivity and optical conductivity spectra around the lowest energy phonon modes of CaBaCo$_4$O$_7$, CaBaFe$_2$Co$_2$O$_7$, and CaBaFe$_4$O$_7$  for light polarization $\mathbf{E}^\omega\parallel{z}$.
The reflectivity spectra over the whole photon energy range covered by our experiment for both $\mathbf{E}^\omega\parallel{z}$ and $\mathbf{E}^\omega\perp{z}$ are presented in the supplement~\cite{Kocsis2020PRB2SM}.
The phonon spectra of CaBaCo$_4$O$_7$ and CaBaFe$_4$O$_7$ (see Figs.~\ref{Swed_01}(a,d), S3 and \ref{Swed_01}(c,f), S4, respectively) change markedly with temperature.
The resonances are narrow at low temperatures and get significantly broader above the magnetic ordering temperature.
Contrary to the pristine compounds, the phonon modes of CaBaFe$_2$Co$_2$O$_7$ depend weakly on the temperature and show no anomaly at $T_{\rm N}$, as shown in Figs.~\ref{Swed_01}(b,e), and S5.

\begin{center}
    \begin{figure}
 
    \includegraphics[width=8.5truecm]{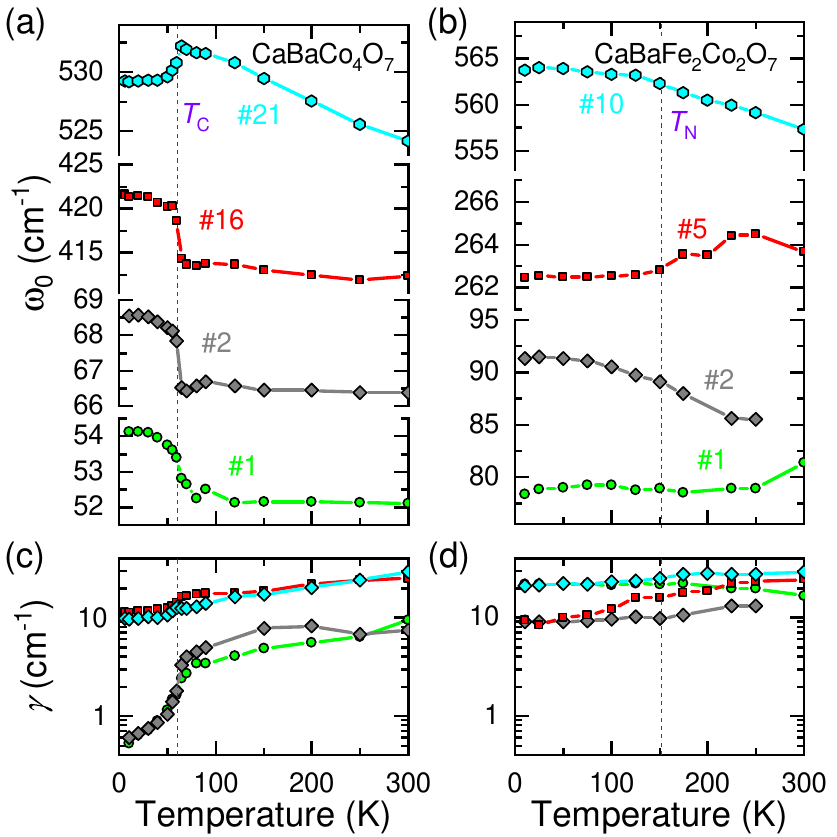}
    \caption{(a,b) Temperature dependence of the fitted phonon frequencies ($\omega_0$) and (c,d) damping rates ($\gamma$) in CaBaCo$_4$O$_7$ and CaBaFe$_2$Co$_2$O$_7$, respectively. The strong coupling between magnetic and elastic properties in CaBaCo$_4$O$_7$ is demonstrated by the changes in $\omega_0$ and $\gamma$ around the magnetic phase transition ($T_{\rm C}$), indicated by dashed lines.}
    \label{Swed_02}
    \end{figure}
    \end{center}

In Fig.~\ref{Swed_02}, we compare the temperature dependence of the phonon parameters, frequency ($\omega_{0,j}$) and damping rate ($\gamma_j$) in CaBaCo$_4$O$_7$ and CaBaFe$_2$Co$_2$O$_7$ for selected, well-separated phonon modes.
In the orthorhombic CaBaCo$_4$O$_7$ and CaBaFe$_4$O$_7$, the phonon modes are non-degenerate already at room temperature, and we did not resolve new modes below the magnetic phase transition temperatures.
However, in both compounds the phonon frequencies change abruptly at the onset of the ferrimagnetic phase transitions.
As an example, the magnitude of phonon energy shift becomes as large as $\Delta\omega_0/\omega_0\sim$4\,\% for modes \#1 and \#2 in CaBaCo$_4$O$_7$, shown in Fig.~\ref{Swed_02}(a).
This is significantly higher than $\Delta\omega_0/\omega_0\sim$1\,\%, the highest value observed in other multiferroics~\cite{Laverdiere2006PRB,Basistyy2014PRB,Reschke2020} and in magnets with strong spin-phonon coupling~\cite{Wakamura1988JAP,Ulrich2015PRL}.
This indicates an extremely strong spin-lattice coupling~\cite{Baltensperger1968,Baltensperger1970,souchkov:027203,Fennie2006PRL}, which agrees with recent experiments demonstrating giant magnetostriction~\cite{Chai2021PRB}.
In CaBaFe$_2$Co$_2$O$_7$, however, the phonon frequencies change slightly with the temperature and we could not resolve any splitting of the phonon modes (see Fig.~S5 and S8).

The most remarkable changes in the infrared spectra of CaBaCo$_4$O$_7$ and CaBaFe$_4$O$_7$ are the drastic increase in the damping rates of all phonon modes as warmed above the ferrimagnetic phase transitions, see Fig.~\ref{Swed_02} and S8, respectively.
Modes \#1 and \#2 of CaBaCo$_4$O$_7$ well exemplify this tendency:
At $T$=10\,K the damping rates of these modes are as low as 0.5\,cm$^{-1}$.
Such sharp phonons with $\gamma$/$\omega_0 <$ 1\,$\%$ are unusual in condensed matter systems, and only observed in non-magnetic molecular crystals~\cite{Dlott1986,Foggi1992PhononRI,Fujioka2009PRL,Fujioka2009PRB}.
However, in the vicinity of $T_{\rm C}$ the phonon lifetime decreases, \textit{i.e.} the damping rate grows by an order of magnitude indicating a strong scattering of phonons by spin-fluctuations.
In the paramagnetic phase, $\gamma$ keeps increasing and at room temperature the phonon modes are strongly damped with $\gamma$/$\omega_0$ ratios exceeding 10\,\%.
The strong temperature dependence of the damping rates away from $T_{\rm C}$, besides the strong spin-lattice coupling, suggests strong lattice anharmonicity~\cite{Klemens1966PR,Balkanski1983PRB}.
The damping rates of modes $\#$16 and $\#$21, and those of CaBaFe$_4$O$_7$ (see Fig. S8) follow similar temperature dependence with pronounced change at the ferrimagnetic phase transitions.
In contrast, the damping rates in CaBaFe$_2$Co$_2$O$_7$ show weak temperature dependence and no anomalies at $T_{\rm N}$.

\begin{center}
    \begin{figure}
 
    \includegraphics[width=8.5truecm]{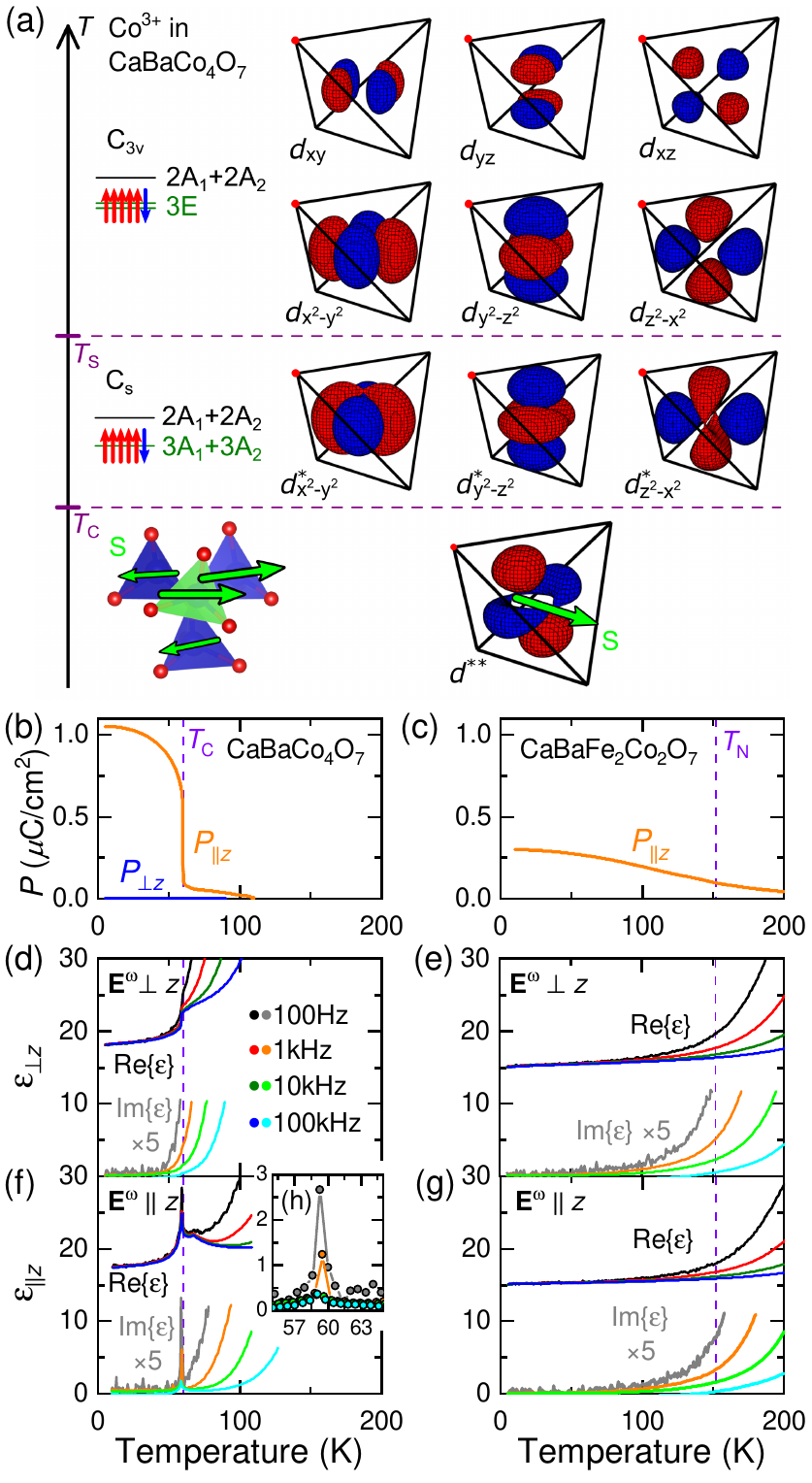}
    \caption{(a) Schematics of the ground state multiplet structure of Jahn-Teller active Co$^{3+}$ ion in CaBaCo$_4$O$_7$. The Jahn-Teller active Fe$^{2+}$ ion in CaBaFe$_4$O$_7$ has the same multiplet structure.
    The magnetic ions in the tetrahedral environment (T$_{d}$) have the orbital-degenerate $^5E$ ground state, which is preserved by the spin orbit interaction.    
    At high temperature ($T_{\rm S}<T$), the oxygen environment is distorted to the polar C$_{3v}$ symmetry, but the orbital degeneracy is preserved by the $E$ ground states $\lbrace{d}_{x^2-y^2},d_{xy}\rbrace$.
    The trigonal to orthorombic distortion decreases the local symmetry to monoclinic $C_s$ ($T_{\rm C}<T<T_{\rm S}$), releases the orbital degeneracy ($\lbrace{d}^*_{x^2-y^2}\rbrace$), and deforms the orbitals.
    The ordering to the ferrimagnetic magnetic state ($T<T_{\rm C}$) further distorts the orbitals and selects only one ($d^{**}$).
    Temperature dependence of the (b,c) pyroelectric polarization and (d-g) dielectric constant of CaBaCo$_4$O$_7$ and CaBaFe$_2$Co$_2$O$_7$, respectively. The (h) inset shows the peak in the imaginary part of $\epsilon_{\parallel{z}}$ at $T_{\rm C}$.}
    \label{Swed_03}
    \end{figure}
    \end{center}

As demonstrated in Fig.~\ref{Swed_03} and S9, the emergence of magnetic order strongly influences the pyroelectric polarization and the low-frequency dielectric response of CaBaCo$_4$O$_7$.
We observed large magnetic-order-induced polarization change for $P\parallel{z}$ in agreement with former results~\cite{Caignaert2013,Johnson2014} and negligible for $P\perp{z}$~\cite{Iwamoto2012}.
The real part of the dielectric constants, both $\epsilon_{\perp{z}}$ and $\epsilon_{\parallel{z}}$, exhibit a step-like change when crossing $T_{\rm C}$ [see Fig.~\ref{Swed_03}(d,f)], with similar magnitude to that of in DyMn$_2$O$_5$ showing colossal magnetodielectric effect~\cite{Sushkov2007}.
Since the step height is independent of frequency between $10^2$ and $10^5$\,Hz, and observed for both $\epsilon_{\perp{z}}$ and $\epsilon_{\parallel{z}}$, the drop in the static dielectric function is related to the sudden changes in the phonon resonances.
In addition to the step-edge in the real part, both the real and the imaginary parts of $\epsilon_{\parallel{z}}$ have a peak at the close vicinity of $T_{\rm C}$.
The frequency dependence and the related finite dissipation indicate electric dipoles with low-frequency dynamics and strong scattering.
The peak shape in the real part suggests that the magnetic fluctuations can couple to electric dipoles and contribute to the phonon scattering~\cite{Lawes2003,Lawes2009}.
Toward higher temperatures, the dielectric constants increase, not due to the change of phonon frequency but due to the decrease  of the resistivity caused by the thermally activated carriers, as shown in Fig.~S2. 
Although CaBaFe$_2$Co$_2$O$_7$ has a similar pyroelectric crystal structure and a relatively high $T_{\rm N}$, its polarization is not affected by the antiferromagnetic order, as displayed in Fig.~\ref{Swed_03}(c).
The dielectric properties of this compound show a smooth variation on temperature in accordance with the phonon spectrum.

We now discuss the enhanced scattering of phonons by spin fluctuations and the origin of the strong anomaly in the dielectric constant observed only in the pristine Swedenborgites, CaBaCo$_4$O$_7$ and CaBaFe$_4$O$_7$.
Remarkably, such a large drop of the phonon damping rate induced by magnetic ordering is rare.
Only minor changes in the damping rate have been detected in emblematic multiferroics including manganites $R$MnO$_3$ ($R$ = Ho, Y)~\cite{Litvinchuk2004,Zaghrioui2008PRB}, TbMnO$_3$~\cite{Schleck2010PRB}, $R$Mn$_2$O$_5$ ($R$ = Tb, Eu, Dy, Bi)~\cite{Aguilar2006,Garciaflores2006PRB}, delafossite CuFeO$_2$~\cite{Aktas2011} or Ni$_3$V$_2$O$_8$~\cite{Vergara2009}.
Although several different mechanisms are responsible for the spin-lattice coupling in these materials, ranging from exchange striction~\cite{Jia2007,Matsumoto2017JPSJ}, inverse Dzyaloshinskii-Moriya interaction~\cite{Katsura2005,Sergienko2006} to on-site anisotropy term~\cite{Arima2007}, none of them results in such a strong magnetic-order-induced change of phonon life-time.
We note that charge fluctuations are frozen in the studied Swedenborgites as indicated by the large dc resistivity and the corresponding few-100\,meV optical charge gap (see Fig.~S2 and S7), thus, these cannot modify the spin-lattice interaction. 
Instead, we argue that low-energy fluctuations of the orbital degrees of freedom open a new channel and mediate a more efficient spin-lattice interaction in CaBaCo$_4$O$_7$ and CaBaFe$_4$O$_7$ since orbitals can strongly interact with both spin fluctuations and phonons.
This may lead to considerable broadening of phonon modes when the ordered state becomes paramagnetic as demonstrated in LaTiO$_3$~\cite{Ulrich2015PRL}.
It is instructive to compare the case of Swedenborgites to that of hexagonal manganites.
Although both class of compounds crystallize in a polar structure with geometric frustration, the phonons are scattered strongly by spin fluctuations exclusively in the Swedenborgites.
In hexagonal mangnites, Mn$^{3+}$ ions sit in a trigonal bipyramid, thus, they have $S=2$ spins just like tetrahedrally coordinated Co$^{3+}$ and Fe$^{2+}$ ions, however, they are not Jahn-Teller active and their orbital singlet ground state is well separated from other 3d states~\cite{Degenhardt2001,Lee2008a}.
This fact also suggests that presence of orbital degrees of freedom allows the unusually strong spin-lattice coupling in Swedenborgites.
Finally, we mention that a recent study of infrared phonons in Fe$_2$Mo$_3$O$_8$ shows similar enhancement of the damping rate across its antiferromagnetic phase transition~\cite{Reschke2020}.

In CaBaCo$_4$O$_7$ and CaBaFe$_4$O$_7$, both the tetrahedrally coordinated Co$^{3+}$ and Fe$^{2+}$ ions possess the orbital-degenerate $^5E$ ground state multiplet as shown in the inset of Fig.~\ref{Swed_01}.
The orbital degeneracy is released by the trigonal to orthorhombic phase transition at $T_{\rm S}$, as illustrated in Fig.~\ref{Swed_03}(a).
The symmetry of the surrounding oxygen ligands is reduced to monoclinic, the $d_{x^2-y^2}$ and $d_{xy}$ orbitals are separated by a small energy gap,  and mixed with $d_{3z^2-r^2}$ orbitals~\cite{Chatterjee2011,Hollmann2011}.
Since these strongly fluctuating low-symmetry orbitals can efficiently couple to the lattice, the phonons strongly scatter on this hybridized ground state in the paramagnetic phase.
As the magnetic order develops, the second-order spin-orbit interaction can further polarize the orbitals, as an example spins along the $y$ axis favours the $d_{z^2-x^2}$ orbital~\cite{Ohtani2010,Nii2012PRB}.
The magnetic order in CaBaCo$_4$O$_7$ selects the same orbital shape at each Co$^{3+}$ site and consequently reduces the fluctuations.
According to this scenario, the quenching of the orbitals at $T_{\rm C}$ strongly influences the lattice as well~\cite{Caignaert2010}, which explains the exceptionally large magnetostriction, magnetic-order-induced polarization, and change in the dielectric response in CaBaCo$_4$O$_7$ and CaBaFe$_4$O$_7$.
The on-site anisotropy as well as the orbital dependence of the exchange interactions (Kugel-Khomski\u{\i}-type interaction) may equally play an important role in the enhanced spin-phonon coupling, however, our experiment is sensitive only to the $\Gamma$-point lattice vibrations, thus it cannot distinguish between these mechanisms.
On one hand, the orbitals may affect the bond orientation dependence of the exchange and its bond-length variation.
On the other hand, they may distort the local environment and spins drive a distortion of the local coordination.
This question may be addressed by studying the momentum dependence of the phonon dispersion and lifetime in a scattering experiment.
As the magnetic ions in CaBaFe$_2$Co$_2$O$_7$ have exclusively orbital-singlet ground states, the magnetic order has no effect on the orbitals and the absence of orbital degrees of freedom diminishes the spin-lattice coupling.
Furthermore, orbital degeneracy can be the driving force behind the phonon anomalies in Fe$_2$Mo$_3$O$_8$~\cite{Reschke2020}, as it contains tetrahedrally coordinated Fe$^{2+}$ ions with orbital degrees of freedom, which suggests that the orbitals can enhance magetoelastic and magnetoelectric couplings not only in Swedenborgites, but also in broader classes of multiferroics.
This idea is further supported by the effect of Ni-doping in CaBaCo$_4$O$_7$, where the substitution of orbital singlet Co$^{2+}$ to Ni$^{2+}$ ions with orbital degeneracy leads to further enhancement of the ME effect~\cite{Gen2022PRB}.
Although precise theoretical description of these materials is challenging, we believe these findings will motivate further experimental and theoretical research.

\section*{Acknowledgments}
The authors are grateful to Karlo Penc for fruitful discussions, and to Akiko Kikkawa and Markus Kriener for the technical assistance.
V.K. was supported by the Alexander von Humboldt Foundation.
This work was supported by the Hungarian National Research, Development and Innovation Office – NKFIH grants FK~135003 and the bilateral program of the Estonian and Hungarian Academies of Sciences under the contract NKM~2021-24, and by the Estonian Research Council grant PRG736, institutional research funding IUT23-3 of the Estonian Ministry of Education and Research and the European Regional Development Fund project TK134.
Illustration of the structural unit cell was created using the software \texttt{VESTA}\cite{Momma2011}.


\bibliographystyle{apsrev4-2}      

%

\cleardoublepage

\newpage
\newpage

\renewcommand{\thefigure}{S\arabic{figure}}
\renewcommand{\theequation}{S\arabic{equation}}
\renewcommand{\thetable}{S\arabic{table}}
\setcounter{figure}{0}
\cleardoublepage

\begin{center}
\textbf{Supplementary Material}
\end{center}

\section{Additional experimental data}

    \begin{center}
    \begin{figure}[h]
 
    \includegraphics[width=8.0truecm]{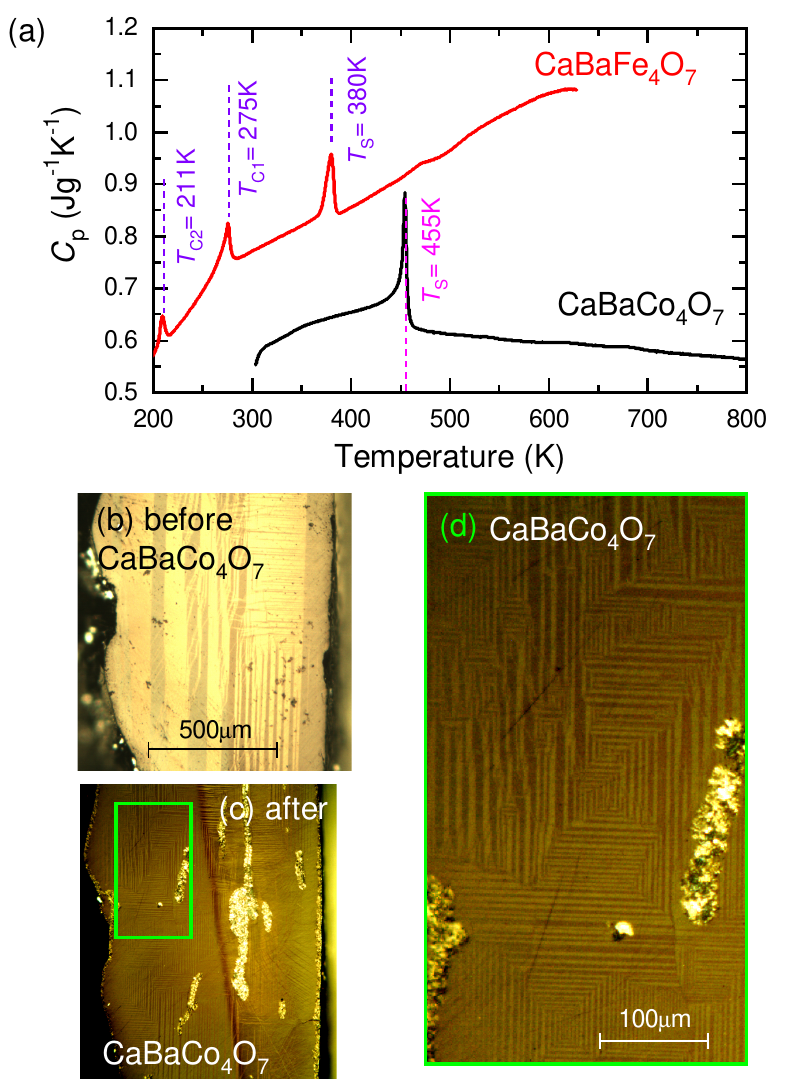}
    \caption{(Color online) (a) Specific heat of CaBaCo$_4$O$_7$ and CaBaFe$_4$O$_7$ measured for warming runs.
    Specific heat data of CaBaFe$_4$O$_7$ is reproduced after Ref.~\onlinecite{Kocsis2016PRB}.
    (b-d) Optical microscopy images (b) before and (c) after the specific heat measurements on CaBaCo$_4$O$_7$. Dark and light contrasted regions correspond to the orthorombic domains. CaBaCo$_4$O$_7$ shows strong twinning on the microscopic scale. (c) Following the specific heat measurements, the orthorombic domains rearrange in a meander-like pattern. Panel (d) shows a magnified region in panel (c).}
    \label{Swed_S01}
    \end{figure}
    \end{center}

Figure~\ref{Swed_S01}(a) shows the specific heat of CaBaCo$_4$O$_7$ and CaBaFe$_4$O$_7$ measured in the warming runs.
The orthorombic to trigonal phase transition temperatures are $T_{\rm S}$=455\,K for CaBaCo$_4$O$_7$ and $T_{\rm S}$=380\,K for CaBaFe$_4$O$_7$.
Above $T_{\rm S}$, neither materials show any further phase transitions.
Figures~\ref{Swed_S01}(b-d) show optical microscopy images of CaBaCo$_4$O$_7$ before and after a high-temperature heat treatment procedure.
The sample was heated to $T$=600\,K for 4\,h in air, then quenched to room temperature.
The microscope images were made in the so-called crossed Nicholson configuration; dark and light contrasted regions correspond to the orthorombic domains.
CaBaCo$_4$O$_7$ shows strong twinning on the microscopic scale. 
After the heat treatment procedure in Fig.~\ref{Swed_S01}(c,d), the orthorombic domains rearrange in a meander-like pattern.

\begin{center}
    \begin{figure}
 
    \includegraphics[width=8.0truecm]{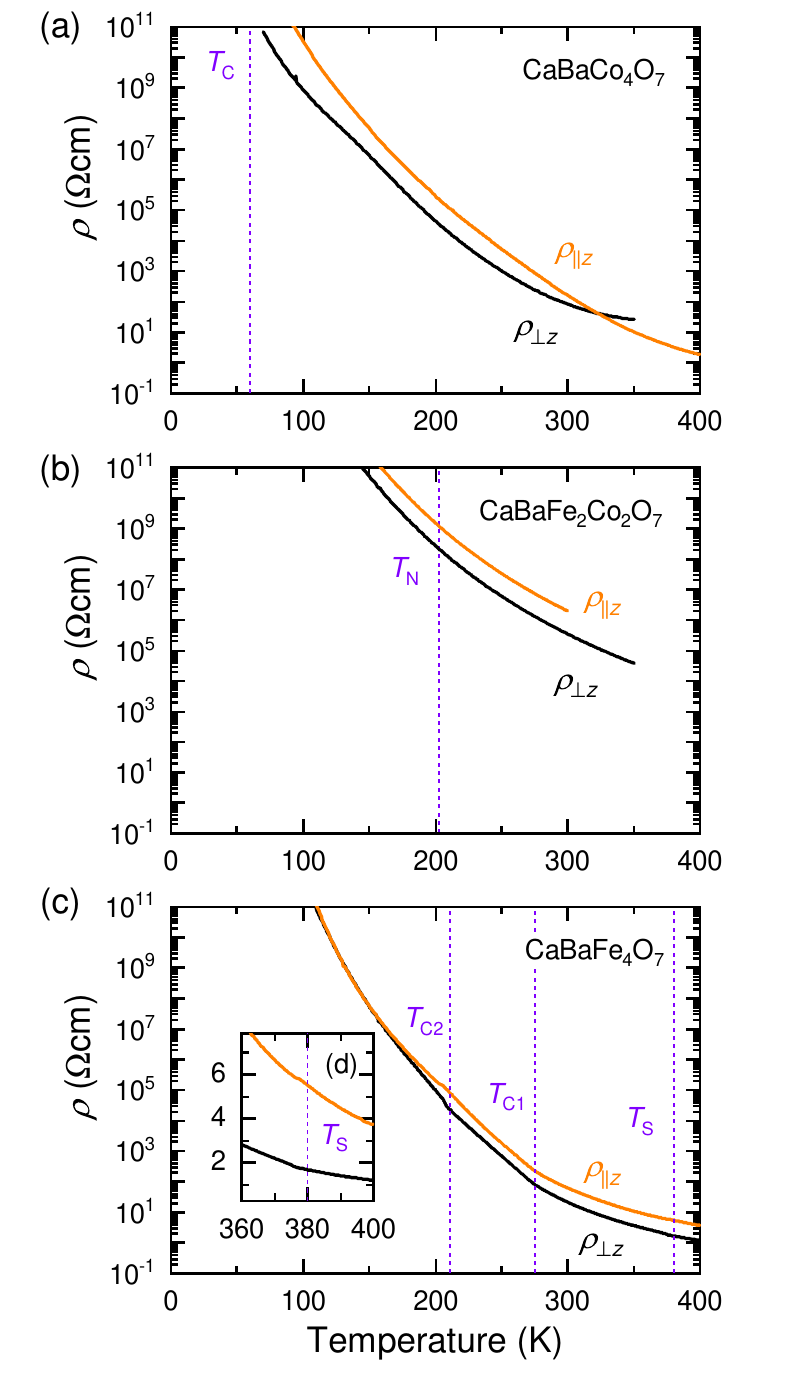}
    \caption{(Color online) Temperature dependence of the resistivity of (a) CaBaCo$_4$O$_7$, (b) CaBaFe$_2$Co$_2$O$_7$, and (c) CaBaFe$_4$O$_7$, measured with currents parallel ($\rho_{\parallel{z}}$) and perpendicular to the $z$-axis ($\rho_{\parallel{z}}$). Panel (d) shows the resistivity of CaBaFe$_4$O$_7$ at $T_{\rm S}$. Note, that CaBaFe$_4$O$_7$ shows very little change in the resistivity, \textit{i.e.} there is definitely no charge order-disorder type of transition accompanying the structural phase transition.}
    \label{Swed_S02}
    \end{figure}
    \end{center}

Figure~\ref{Swed_S02} shows the temperature dependence of the resistivity ($\rho$) in CaBaCo$_4$O$_7$, CaBaFe$_2$Co$_2$O$_7$, and CaBaFe$_4$O$_7$.
The resistivity was measured with currents parallel ($\rho_{\parallel{z}}$) and perpendicular to the $z$ axis ($\rho_{\perp{z}}$).
The resistivity data of CaBaFe$_4$O$_7$ shows only a very subtle anomaly at $T_{\rm S}$, and has semiconductor-like temperature dependence both below and above $T_{\rm S}$.
The absence of strong anomalies in the specific heat and resistivity data at temperatures above $T_{\rm S}$ in Figs.~\ref{Swed_S01} and \ref{Swed_S02} implies that the charge ordered state is not melted up to the decomposition temperatures in CaBaFe$_4$O$_7$.

Figures~\ref{Swed_S03}, \ref{Swed_S04}, \ref{Swed_S05}, and \ref{Swed_S06} show the reflectivity and optical conductivity spectra of CaBaCo$_4$O$_7$, CaBaFe$_4$O$_7$, CaBaFe$_2$Co$_2$O$_7$, and YBaCo$_3$AlO$_7$ at selected temperatures.
In each figure, panels (a,c) and (b,d) correspond to measurements with $\mathbf{E}^\omega\perp{z}$ and $\mathbf{E}^\omega\parallel{z}$, respectively.
YBaCo$_3$AlO$_7$ is a spin-glass ($T_{\rm f}$=17\,K), has the hexagonal P6$_3$mc structure~\cite{Valldor2008}, and it hosts exclusively orbital-singlet Co$^{2+}$ ions.
Comparison of YBaCo$_3$AlO$_7$ to CaBaFe$_2$Co$_2$O$_7$ and to the ferrimagnetic compounds helped us to examine the role of long-range order.
Similarly to the solid solution CaBaFe$_2$Co$_2$O$_7$, the phonons of YBaCo$_3$AlO$_7$ in Fig.~\ref{Swed_S05} show only weak temperature dependence.

The optical conductivity spectra were calculated from the reflectivity data using the Kramers-Kronig transformation.
The low-energy part of the measured reflectivity spectra was extrapolated to zero photon energy as a constant value.
Figure~\ref{Swed_S07} shows the UV and hard-UV optical reflectivity and optical conductivity of CaBaCo$_4$O$_7$, CaBaFe$_2$Co$_2$O$_7$, and CaBaFe$_4$O$_7$, which was used as a high-energy extension for the Kramers-Kronig transformation.
Spectra above $k$=10$^6$\,cm$^{-1}$ were assumed to follow the free electron model.

Figure~\ref{Swed_S08} summarizes the temperature dependence of the fitted phonon frequencies ($\omega_0$), oscillator strengths ($S$), and damping rates ($\gamma$) in CaBaCo$_4$O$_7$, CaBaFe$_2$Co$_2$O$_7$, and CaBaFe$_4$O$_7$.
Data shown in Fig.~\ref{Swed_S08}(a,b,g,h) are the same as those in Fig.~2. 
At the magnetic phase transitions, the phonon frequencies and damping rates show remarkable changes in the pristine compounds.
In CaBaFe$_2$Co$_2$O$_7$, none of the phonon parameters show change in the vicinity of $T_{\rm N}$.
We detect the appearance of no new phonon modes, which is supported by the temperature dependence of $S$ in \ref{Swed_S08}(d,e,f), which show changes only around the magnetic phase transitions.

Figures~\ref{Swed_S09}(a-c) show the real and imaginary parts of the dielectric constants ($\epsilon$) measured in CaBaCo$_4$O$_7$, CaBaFe$_2$Co$_2$O$_7$, and CaBaFe$_4$O$_7$, respectively for $\mathbf{E}^\omega\perp{z}$ in the upper panels and $\mathbf{E}^\omega\parallel{z}$ in the lower panels.
The real and imaginary parts of the ac magnetic susceptibility (ac-$\chi$) measured in  CaBaCo$_4$O$_7$, CaBaFe$_2$Co$_2$O$_7$, and CaBaFe$_4$O$_7$ for $\mathbf{H}^\omega\perp{z}$ and $\mathbf{H}^\omega\parallel{z}$ are shown in Figs.~\ref{Swed_S09}(d), \ref{Swed_S09}(e), and \ref{Swed_S09}(f), respectively.
The magnitude of the oscillating magnetic field was $\delta\mathbf{H}^\omega$=5\,Oe.
The ac-$\chi$ measurements were performed in the absence of static $H$ field, except for the lower panel of Fig.~\ref{Swed_S09}(f), where a moderate $H$ = 3\,kOe static field was applied.
In CaBaCo$_4$O$_7$, the real part of $\epsilon_{\perp{z}}$ has a step like jump, while the imaginary part rapidly increases above $T_{\rm C}$.
The real part of $\epsilon_{\parallel{z}}$ has a double peak structure (strongest peak at $T_{\rm C}$), and the imaginary part has a single peak at $T_{\rm C}$.
Above $T_{\rm C}$, all components of the dielectric constants show strong frequency dependence and anisotropy, \textit{i.e.} $\epsilon_{\parallel{z}}$ increases more rapidly with temperature than $\epsilon_{\perp{z}}$.
However, for low frequencies these features are an aggregate of the anisotropic resistivity and the Maxwell-Wagner relaxation caused by Schottky barriers forming at sample electrode interfaces~\cite{Lunkenheimer2002PRB,Lunkenheimer2009EPJ}.
Figures~\ref{Swed_S09}(d) and \ref{Swed_S09}(f) also show the frequency dependence of the imaginary part of the ac-$\chi$ in CaBaCo$_4$O$_7$ and CaBaFe$_4$O$_7$, respectively.
The inset of panel (d), Fig.~\ref{Swed_S09}(g), shows the imaginary part of the ac-$\chi$ measured in CaBaCo$_4$O$_7$ for a magnified region.
The imaginary parts of the ac-$\chi$ both in CaBaCo$_4$O$_7$ and in CaBaFe$_4$O$_7$ has asymmetric peaks around $T_{\rm C}$ and $T_{\rm C2}$, respectively, which means increased dissipation on the magnetic domain walls at low-frequencies (below 1\,kHz).
In CaBaCo$_4$O$_7$, the broad symmetric peak at $T_{\rm C}$ in the real part of $\chi_{\perp{z}}$ is accompanied by an asymmetric peak in the imaginary part, and a step in $\rm{Re}\lbrace\chi_{\parallel{z}}\rbrace$.
Frequency dependence of the magnetic $\rm{Im}\lbrace\chi_{\perp{z}}\rbrace$ resembles to that of the dielectric $\rm{Im}\lbrace\epsilon_{\parallel{z}}\rbrace$, however at much lower frequencies.
In contrast to the pristine compounds, the antiferromagnetic CaBaFe$_2$Co$_2$O$_7$ has no features in the dielectric constants in Fig.~\ref{Swed_S09}(b), and has only a small peak in the ac magnetic susceptibility in Fig.~\ref{Swed_S09}(e).
Although the ac-$\chi$ has similar features to $\epsilon$, which would suggest a strong connection between magnetic and lattice fluctuations, however, the magnetic fluctuations are at very low frequencies and the strength of the magnetic fluctuations decay quickly towards higher frequencies.
Therefore, magnetic fluctuations alone cannot account for the increased phonon scattering observed in the optical measurements at significantly higher frequencies.
As a conclusion, in CaBaCo$_4$O$_7$ and CaBaFe$_4$O$_7$, the electric and magnetic fluctuations are relevant only in the vicinity of the ferrimagnetic phase transitions, while the magnetic fluctuations are not relevant at optical frequencies.
Therefore, the strong anharmonicity of phonon modes in the pristine compounds are not explained by these fluctuations.

\section{Lattice excitations in Swedenborgites}

Swedenborgites, CaBa$M_4$O$_7$ ($M$=Co, Fe) are built up by alternating layers of triangular and kagom\'e lattices of tetrahedrally coordinated transition metal ions.
In general, these compounds realize a trigonal structure ($P31c$), with $P6_3mc$ as the possible highest symmetry mother-structure.
Note that the space group of the hexagonal manganites is $P6_3cm$.
The irreducible representations of the infrared-active normal modes for the $P6_3mc$ mother structure are:
\begin{equation}
\Gamma_{\rm IR} = 9 A_1 (z) + 12 E_1(xy).
\end{equation}
Namely, for $\mathbf{E}^\omega\parallel{z}$ the reflectivity spectra may show 9 non-degenerate $A_1$ modes, and for $\mathbf{E}^\omega\perp{z}$ 12 doubly degenerate $E_1$ modes.
Note, that YBaCo$_3$AlO$_7$ in Fig.~\ref{Swed_S06} shows 9 modes for $\mathbf{E}^\omega\parallel{z}$.

At high-temperatures (above $T_{\rm S}$), the pristine CaBaFe$_4$O$_7$ and CaBaCo$_4$O$_7$, as well as the solid solution CaBaFe$_2$Co$_2$O$_7$ at all temperatures have the trigonal structure, described by the $P31c$ space group.
The irreducible representations of the infrared-active phonons are:
\begin{equation}
\Gamma_{\rm IR} = 12 A_1 (z) + 25 E(xy).
\end{equation}
Therefore, for $\mathbf{E}^\omega\parallel{z}$ and $\mathbf{E}^\omega\perp{z}$, the reflectivity spectra may show 12 $A_1$ and 25 $E$ modes, respectively.
For $\mathbf{E}^\omega\parallel{z}$ in Fig.~\ref{Swed_S05}(b,d), CaBaFe$_2$Co$_2$O$_7$ shows 12 modes.

\begin{center}
    \begin{figure*}[t!]
 
    \includegraphics[width=17.0truecm]{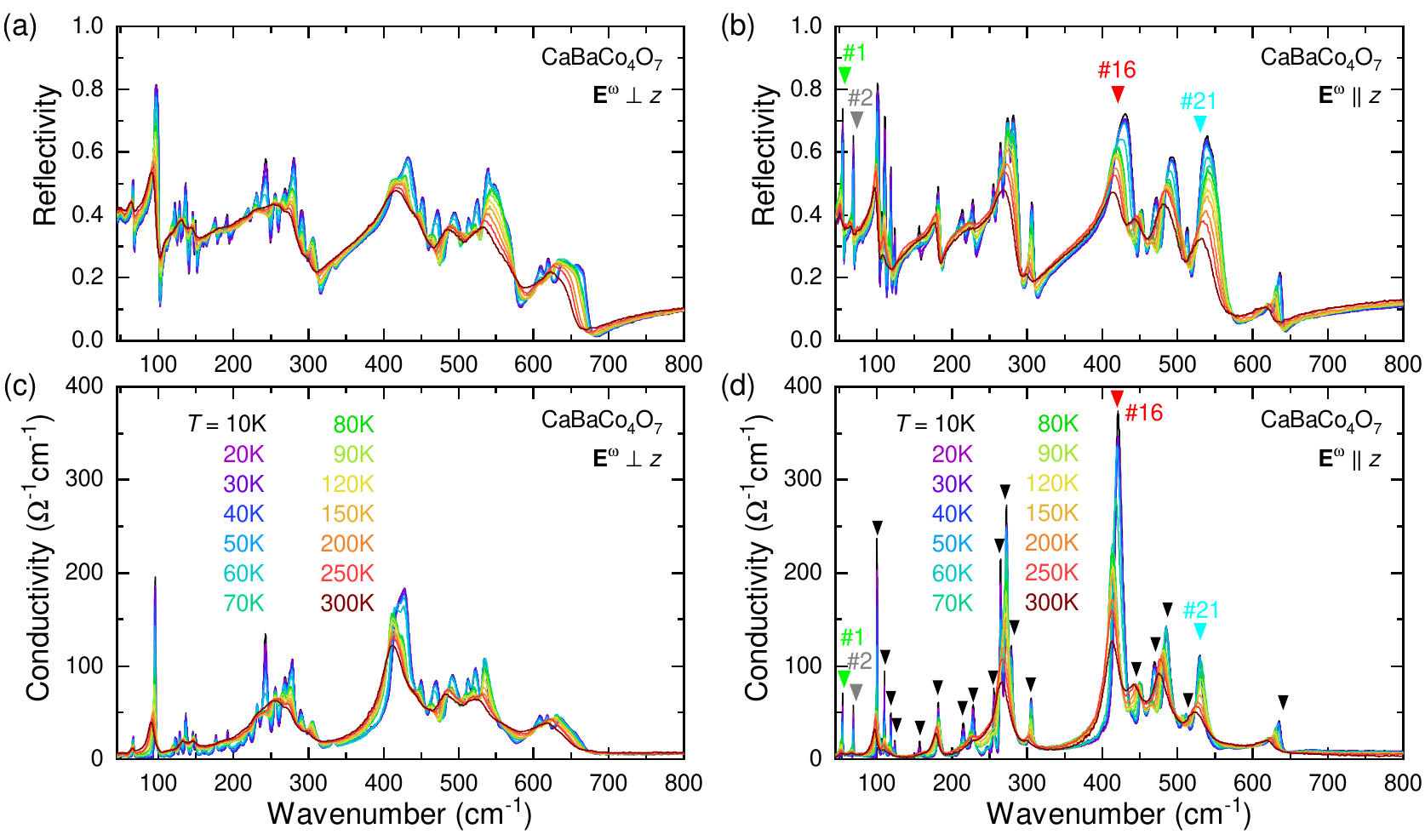}
    \caption{(Color online) Temperature dependence of the (a,b) optical reflectivity and (c,d) calculated optical conductivity spectra of CaBaCo$_4$O$_7$. Panels (a,c) and panels (b,d) show measurements for $\mathbf{E}^\omega\perp{z}$ and $\mathbf{E}^\omega\parallel{z}$, respectively.}
    \label{Swed_S03}
    \end{figure*}
    \end{center}

\begin{center}
    \begin{figure*}[t!]
 
    \includegraphics[width=17.0truecm]{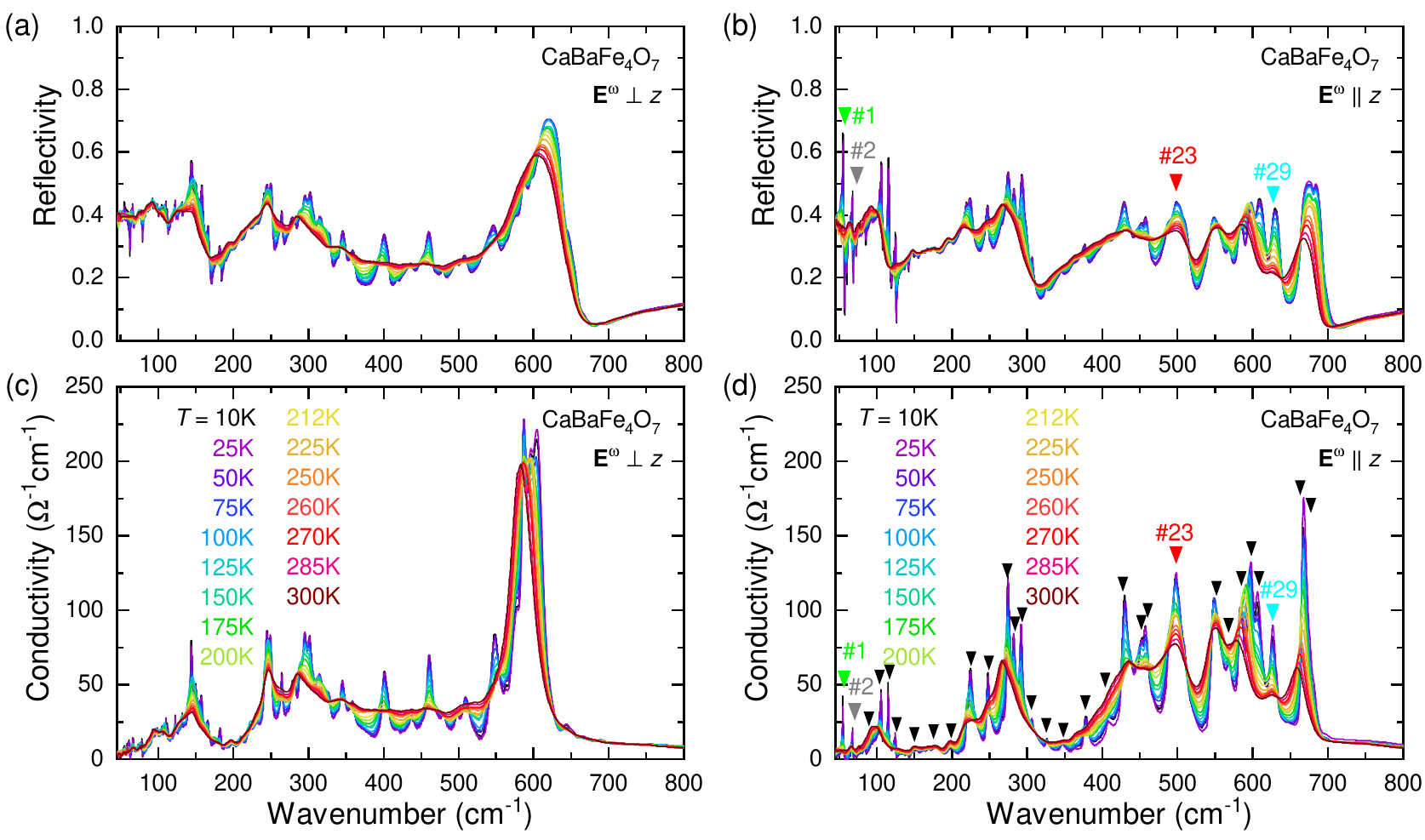}
    \caption{(Color online) Temperature dependence of the (a,b) optical reflectivity and (c,d) calculated optical conductivity spectra of CaBaFe$_4$O$_7$. Panels (a,c) and panels (b,d) show measurements for $\mathbf{E}^\omega\perp{z}$ and $\mathbf{E}^\omega\parallel{z}$, respectively.}
    \label{Swed_S04}
    \end{figure*}
    \end{center}

\begin{center}
    \begin{figure*}[t!]
 
    \includegraphics[width=17.0truecm]{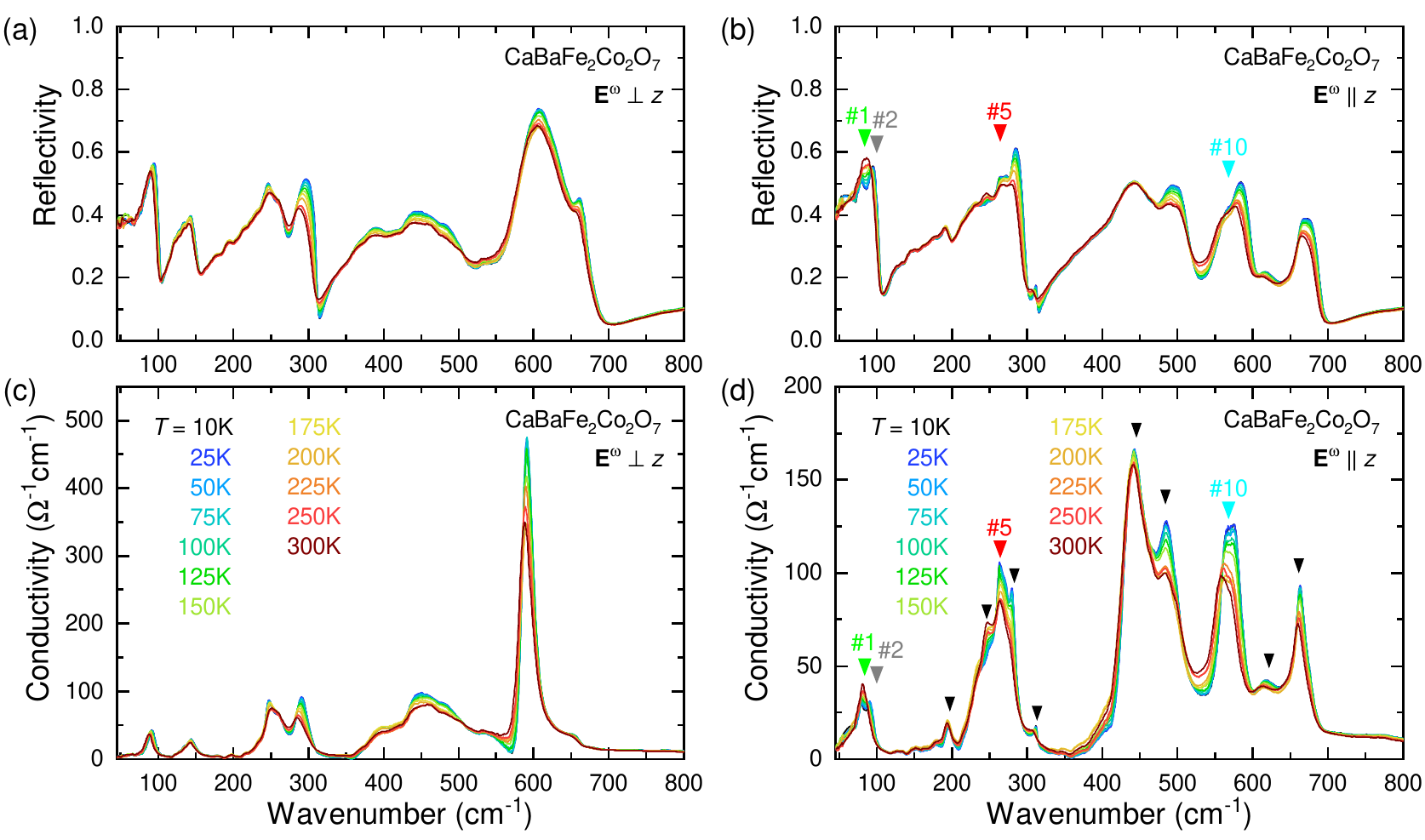}
    \caption{(Color online) Temperature dependence of the (a,b) optical reflectivity and (c,d) calculated optical conductivity spectra of CaBaFe$_2$Co$_2$O$_7$. Panels (a,c) and panels (b,d) show measurements for $\mathbf{E}^\omega\perp{z}$ and $\mathbf{E}^\omega\parallel{z}$, respectively.}
    \label{Swed_S05}
    \end{figure*}
    \end{center}

\begin{center}
    \begin{figure*}[t!]
 
    \includegraphics[width=17.0truecm]{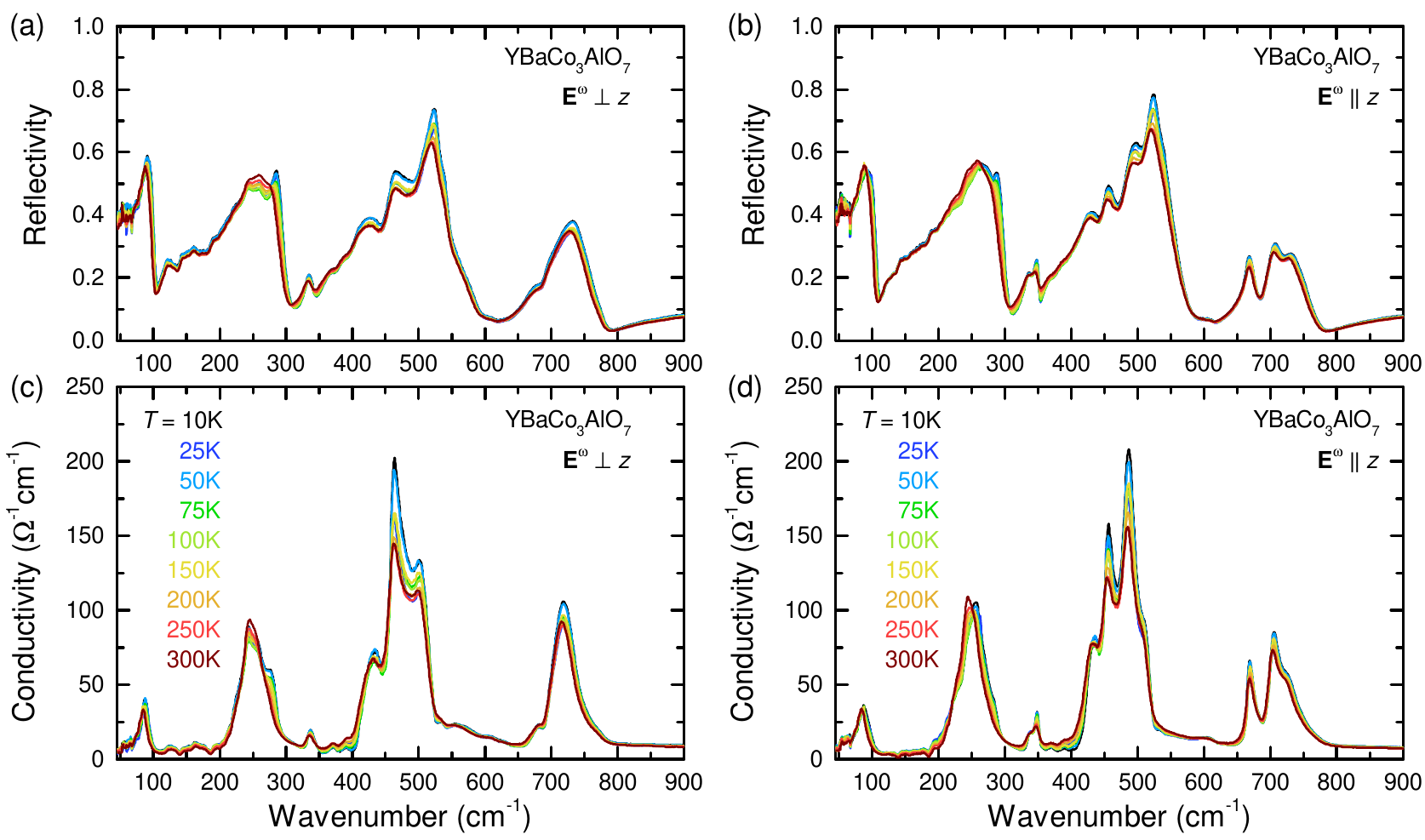}
    \caption{(Color online) Temperature dependence of the (a,b) optical reflectivity and (c,d) calculated optical conductivity spectra of YBaCo$_3$AlO$_7$. Panels (a,c) and panels (b,d) show measurements for $\mathbf{E}^\omega\perp{z}$ and $\mathbf{E}^\omega\parallel{z}$, respectively.}
    \label{Swed_S06}
    \end{figure*}
    \end{center}

\begin{center}
    \begin{figure}[t!]
 
    \includegraphics[width=8.0truecm]{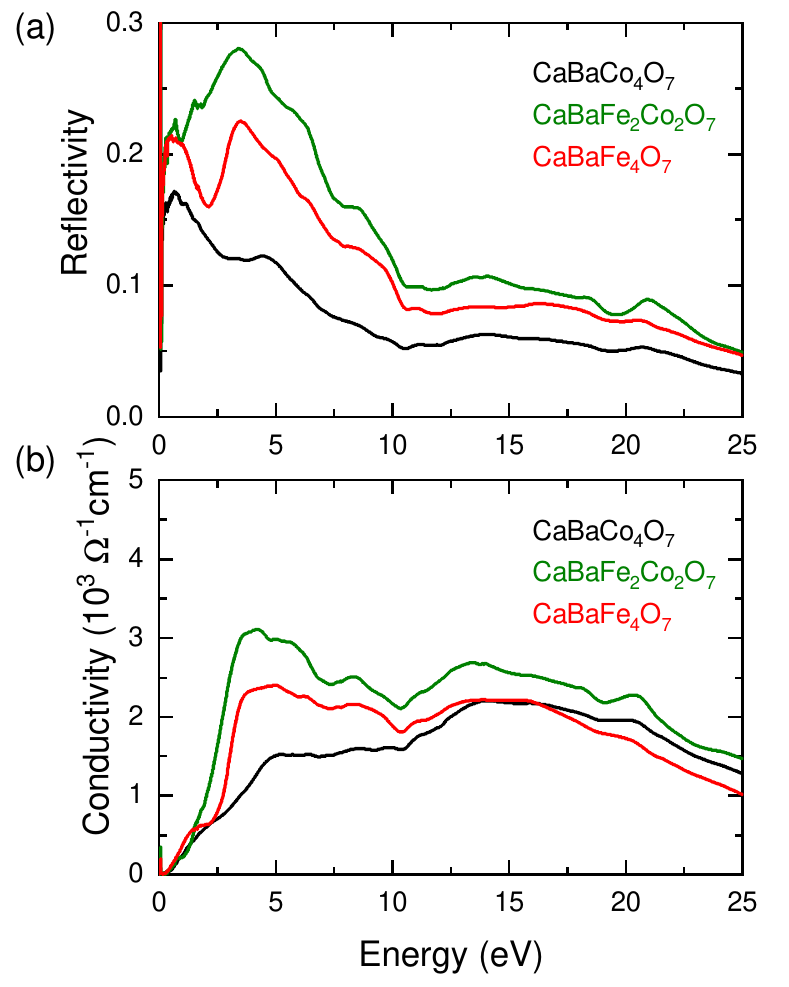}
    \caption{(Color online) (a) Hard UV reflectivity and (b) optical conductivity spectra of CaBaCo$_4$O$_7$, CaBaFe$_2$Co$_2$O$_7$, and CaBaFe$_4$O$_7$ measured at $T$=300\,K.}
    \label{Swed_S07}
    \end{figure}
    \end{center}

\begin{center}
    \begin{figure*}[t!]
 
    \includegraphics[width=15.0truecm]{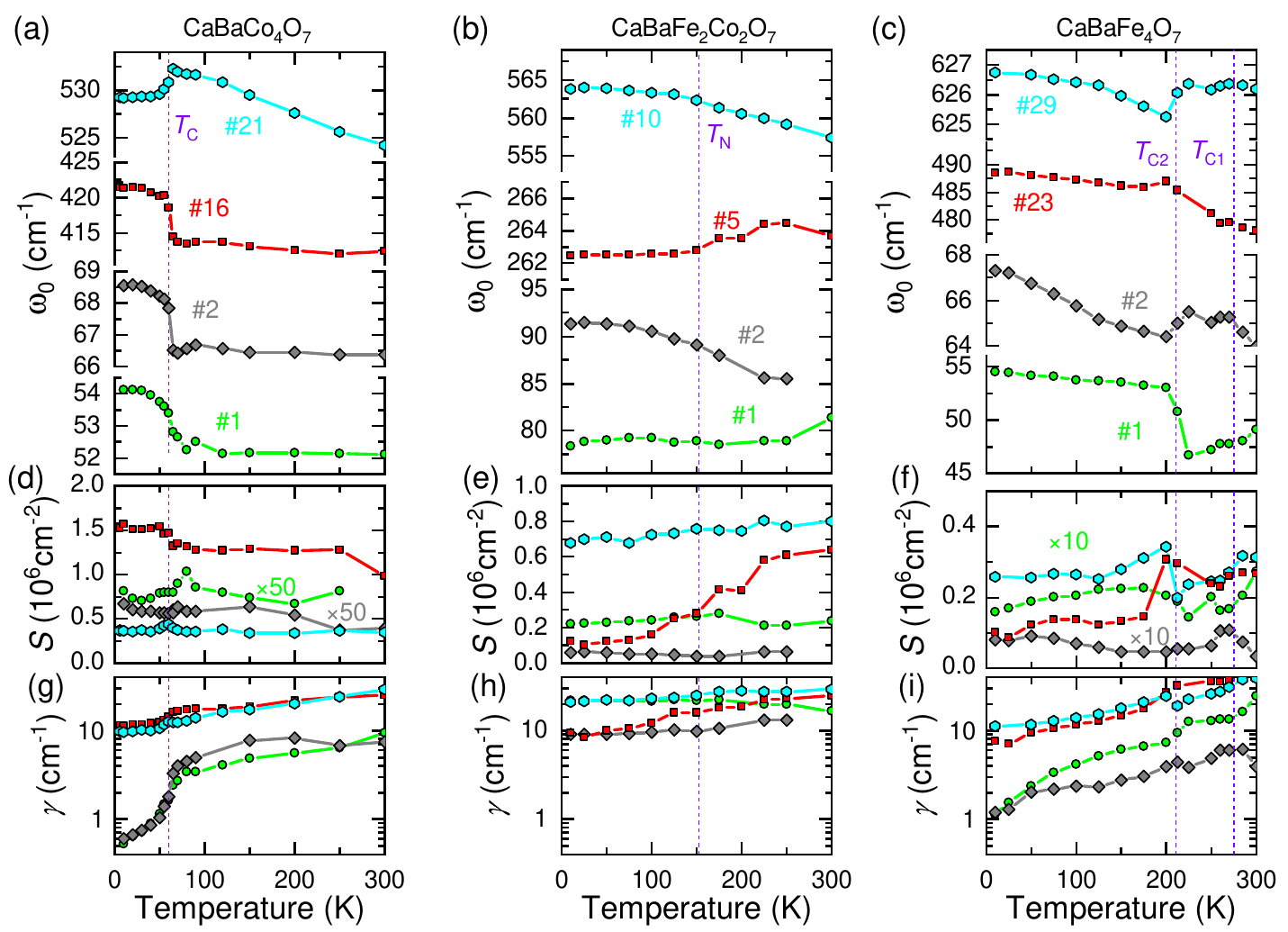}
    \caption{(Color online) Temperature dependence of the fitted (a,b,c) phonon frequencies ($\omega_0$), (d,e,f) oscillator strengths ($S$), and (g,h,i) damping rates ($\gamma$) of CaBaCo$_4$O$_7$, CaBaFe$_2$Co$_2$O$_7$, and CaBaFe$_4$O$_7$, respectively.}
    \label{Swed_S08}
    \end{figure*}
    \end{center}

\begin{center}
    \begin{figure*}[t!]
 
    \includegraphics[width=15.0truecm]{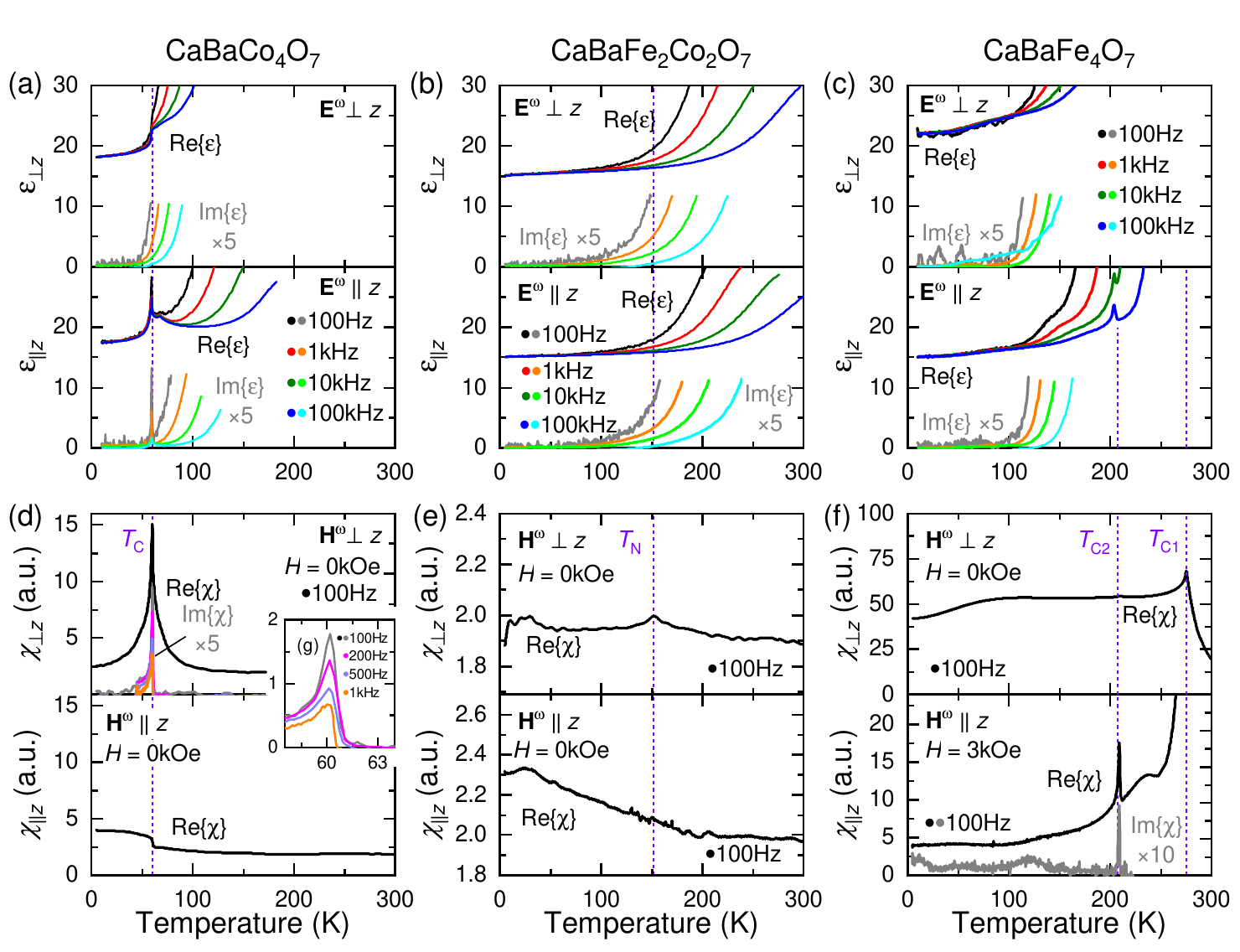}
    \caption{(Color online) Temperature dependence of (a,b,c) the dielectric constant and (d,e,f) the ac magnetic susceptibility of CaBaCo$_4$O$_7$, CaBaFe$_2$Co$_2$O$_7$, and CaBaFe$_4$O$_7$, respectively. Note that the imaginary part of the dielectric constant is multiplied by a factor of 5 for better visibility. (g) The inset shows a magnified region of the imaginary part of the ac-$\chi$ from panel (d).}
    \label{Swed_S09}
    \end{figure*}
    \end{center}

Below $T_{\rm S}$, CaBaCo$_4$O$_7$ and CaBaFe$_4$O$_7$ have the orthorombic $Pbn2_1$ structure and the infrared-active phonons are:
\begin{equation}
\Gamma_{\rm IR} = 39 A_1 (z) + 39 B_1(y) + 39 B_2(x).
\end{equation}
Namely, the reflectivity spectra should show 39 non-degenerate modes for all three polarizations of the electromagnetic radiation.
For $\mathbf{E}^\omega\parallel{z}$ in Figs~\ref{Swed_S03}(b,d) and \ref{Swed_S04}(b,d) we identify 22 and 31 phonon modes for CaBaCo$_4$O$_7$ and CaBaFe$_4$O$_7$, respectively.
The lower number of phonons compared to the expected may come from accidental degenerations or from modes outside the spectral window with $k$=25\,cm$^{-1}$ cutoff energy.
We note that low-energy orbital fluctuations of tetrahedral Fe$^{2+}$ ions can also be active and mix among the phonon excitations.


\end{document}